\documentclass[fleqn,usenatbib]{mnras}
\usepackage{newtxtext,newtxmath}
\usepackage[T1]{fontenc}

\DeclareRobustCommand{\VAN}[3]{#2}
\let\VANthebibliography\thebibliography
\def\thebibliography{\DeclareRobustCommand{\VAN}[3]{##3}\VANthebibliography}

\usepackage{graphicx}	
\usepackage{amsmath}	
\usepackage{comment}
\usepackage{multicol}
\usepackage{xcolor}
\usepackage{xspace}

\newcommand{\ujy}{$\mu$Jy\xspace} 
\newcommand{\ujb}{$\mu$Jy~beam$^{-1}$\xspace}
\newcommand{\whz}{W~Hz$^{-1}$\xspace}
\DeclareUnicodeCharacter{2060}{\nolinebreak}

\title[VLASS stacking of high-redshift AGNs]{Sub-mJy radio emission from high-redshift active galactic nuclei in the footprint of the VLA Sky Survey}

\author[K. Perger et al.]{
Krisztina Perger,$^{1,2}$\thanks{E-mail:perger.krisztina@csfk.org}
S\'andor Frey,$^{1,2,3}$
and Krisztina \'E. Gab\'anyi$^{4,5,1,2}$
\\
$^{1}$ Konkoly Observatory, HUN-REN Research Centre for Astronomy and Earth Sciences, Konkoly Thege Mikl\'{o}s \'{u}t 15-17, Budapest, H-1121, Hungary\\
$^{2}$ CSFK, MTA Centre of Excellence, Konkoly Thege Mikl\'os \'ut 15-17, H-1121 Budapest, Hungary\\
$^{3}$ Institute of Physics and Astronomy, ELTE E\"{o}tv\"{o}s Lor\'{a}nd University, P\'{a}zm\'{a}ny P\'{e}ter s\'{e}t\'{a}ny 1/A, Budapest, H-1117, Hungary\\
$^{4}$ Department of Astronomy, Institute of Physics and Astronomy, ELTE E\"{o}tv\"{o}s Lor\'{a}nd University, P\'{a}zm\'{a}ny P\'{e}ter s\'{e}t\'{a}ny 1/A, Budapest, H-1117, Hungary\\
$^5$ HUN-REN--ELTE Extragalactic Astrophysics Research Group, P\'{a}zm\'{a}ny P\'{e}ter s\'{e}t\'{a}ny 1/A, Budapest, H-111, Hungary
}

\date{Accepted 2023 November 01. Received 2023 November 01; in original form 2023 September 27}

\pubyear{2023}

\begin{document}
\label{firstpage}
\pagerange{\pageref{firstpage}--\pageref{lastpage}}
\maketitle

\begin{abstract}
Using empty-field `Quick Look' images from the first two epochs of the VLA Sky Survey (VLASS) observations, centred on the positions of $\sim3700$ individually radio-non-detected active galactic nuclei (AGNs) at $z\ge4$, we performed image stacking analysis to examine the sub-mJy emission at $3$~GHz. We found characteristic monochromatic radio powers of $P_\mathrm{char}=(2-13) \times 10^{24}$~\whz, 
indicating that AGN-related radio emission is widespread in the sample. The signal-to-noise ratios of the redshift-binned median stacked maps are between $4-6$, and we expect that with the inclusion of the yet to be completed third-epoch VLASS observations, the detection limit defined as signal-to-noise ratio $\mathrm{SNR}\ge6$ could be reached, and the redshift dependence can be determined. To obtain information on the general spectral properties of the faint radio emission in high-redshift AGNs, we confined the sample to $\sim3000$ objects covered by both the VLASS and the Faint Images of the Radio Sky at Twenty-centimeters (FIRST) survey. We found that the flux densities from the median stacked maps show a characteristic spectral index of $\alpha^*=-0.30\pm0.15$, which is in agreement with the median spectral index of the radio-detected $z\ge4$ AGNs from our high-redshift AGN catalogue. The three-band mid-infrared colour--colour diagram based on \textit{Wide-field Infrared Survey Explorer} observations provides further support regarding the AGN contribution to the radio emission in the sub-mJy sample. 
\end{abstract}



\section{Introduction}

There are currently $4124$ spectroscopically identified active galactic nuclei (AGNs) at redshifts $z\ge4$ collected in our continuously updated catalogue of high-redshift AGNs\footnote{\url{https://staff.konkoly.hu/perger.krisztina/catalog.html}} \citep[P17 hereafter,][]{2017FrASS...4....9P}. Around $8$ per cent of them were detected in either of the large sky surveys in radio bands, the Faint Images of the Radio Sky at Twenty-centimeters \citep[FIRST,][]{1997ApJ...475..479W,2015ApJ...801...26H} survey, the NRAO VLA Sky Survey \citep[NVSS,][]{1998AJ....115.1693C}, and the VLA Sky Survey \citep[VLASS,][]{2020RNAAS...4..175G,2021ApJS..255...30G}, including $\sim30$ objects for which flux densities are reported in different radio wavebands from individual observations (see Column \textit{`Notes'} in the catalogue). Based on findings of lower-redshift AGNs, the radio-emitting fraction is expected to be around $10$ per cent \citep[e.g.][]{2016ApJ...831..168K}. The somewhat lower fraction of $z\ge4$ AGNs with detected radio emission may be an observational bias caused by the sensitivity limits of the surveys. It was also found that there is a decrease of the radio-loud fraction of AGNs with increasing redshift and decreasing luminosity \citep[e.g.][]{2007ApJ...656..680J,2015AJ....149...61K,2023MNRAS.525.5291L}, with only radio-quiet AGNs expected to be found at the faint end of the optical luminosity function \citep{2011MNRAS.411.1547P}, with the radio-loud fraction estimated to be around or less than 5 per cent. However, others claimed little or no clear evolution of the radio-loud fraction through cosmic time \citep[][]{2016ApJ...829...33Y,2021A&A...656A.137G,2021ApJ...908..124L}.
Indeed, image stacking analyses that allow us to probe the faint radio source population below the detection limit for individual sources revealed the presence of sub-mJy level radio emission in AGNs at all redshifts \citep[e.g.][]{2005MNRAS.360..453W,2007ApJ...654...99W,2008AJ....136.1097H,2015MNRAS.448.2665W,2018MNRAS.477..830H,2019MNRAS.490.2542P,2022A&A...663A.153R}. 

There is still a considerable debate whether there is a dichotomy in the distribution of radio emission in AGNs. On the one hand, there seems ample evidence of a distinction between radio-loud and radio-quiet populations \citep[e.g.][]{1989AJ.....98.1195K,1990MNRAS.244..207M,1999ApJ...511..612G,2007ApJ...654...99W,2016ApJ...831..168K,2023MNRAS.518...39W}. On the other hand, multiple authors claim no clear division between AGNs based on their radio loudness \citep[e.g.][]{2003MNRAS.341..993C,2003MNRAS.346..447C,2012ApJ...759...30B,2013ApJ...768...37C,2019A&A...622A..11G,2021MNRAS.506.5888M}, as the apparent dichotomy is introduced by observational and mathematical biases. 

The radio-loud/radio-quiet dichotomy was also discussed as an effect caused by the different physical origins of the bulk of the detected radio emission. Systems above the $\sim100$~\ujy threshold are dominated by  AGN jets, while the ones with low luminosities and low flux densities are predominantly driven by the ongoing enhanced star formation in the host galaxy, and are radio-quiet AGNs \citep[e.g.][]{2013MNRAS.436.3759B,2017A&A...602A...4D,2017MNRAS.468..946S,2017A&A...602A...2S,2018A&A...620A.192C,2019A&A...622A..11G,2020A&A...642A.125C,2022A&A...663A.153R}. Recent studies also suggest that the observed radio emission most likely originates from the combined effect of star formation and the radio jets of AGNs \citep[e.g.][]{2019MNRAS.490.2542P,2020A&A...642A.125C,2021MNRAS.506.5888M,2022A&ARv..30....6M}.

Large sky surveys in the radio regime like the currently on-going VLASS can provide important insight to that field as well \citep{2020PASP..132c5001L}. In the framework of the VLASS, the entire sky at declinations $\delta \geq -40\degr$ was covered in the S band ($2~\mathrm{GHz} < \nu < 4~\mathrm{GHz}$) with an angular resolution of $2\farcs5$ \citep{2020RNAAS...4..175G,2021ApJS..255...30G}. Gaussian model components were fitted to `flux islands' with peak intensities of $5\sigma$ in VLASS `Quick Look' images  \citep{2021ApJS..255...30G}. The median rms noise levels of $128-145$~\ujb\citep{2021ApJS..255...30G} thus outline a lower limit on components with intensities labeled as detections in the range of of $640-725$~\ujb.

We present the results of our image-stacking analyses carried out using first- and second-epoch VLASS `Quick Look' radio maps centred around the positions of $\sim3700$ radio-non-detected AGNs. For comparison, the stacking analysis was repeated using a slightly smaller sample of $\sim3000$ AGNs, which also fell in the observational footprint of the FIRST survey. For further support, we studied the mid-infrared (MIR) colours of the stacked AGNs and compared them with those of the radio-detected population, utilising \textit{Wide-field Infrared Survey Explorer} \citep[\textit{WISE},][]{2010AJ....140.1868W} data.

We assumed a standard $\Lambda$ cold dark matter (⁠$\Lambda$CDM) cosmological model for the calculations, with $\Omega_\mathrm{m}=0.3$, $\Omega_\Lambda=0.7⁠$, and $H_0=70$~km\,s$^{-1}$\,Mpc$^{-1}$.

\section{Sample selection and the stacking process}

\begin{figure}
 \centering
 \includegraphics[width=\linewidth]{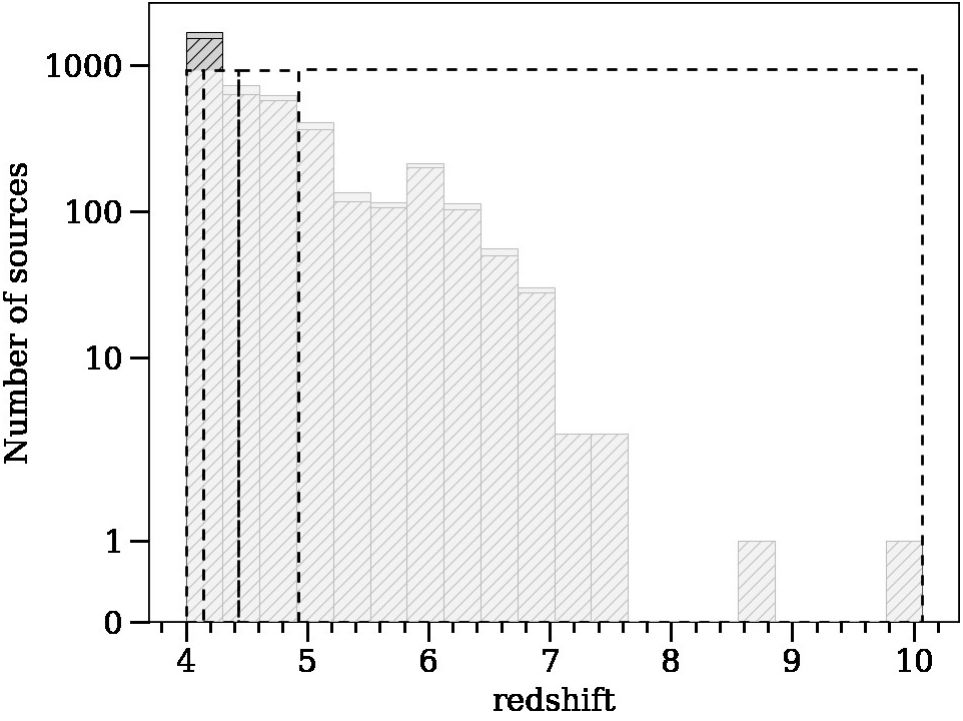}
 \caption{Distribution of all high-redshift AGNs from the P17 catalogue as a function of redshift (light grey) with the VLASS non-detected sources overlaid (hatched). The redshift boundaries for the binned data of Sample~V are indicated with white transparent face and dashed lines.}\label{fig:binning}
\end{figure}

\begin{figure*}
 \centering
 \includegraphics[width=\linewidth]{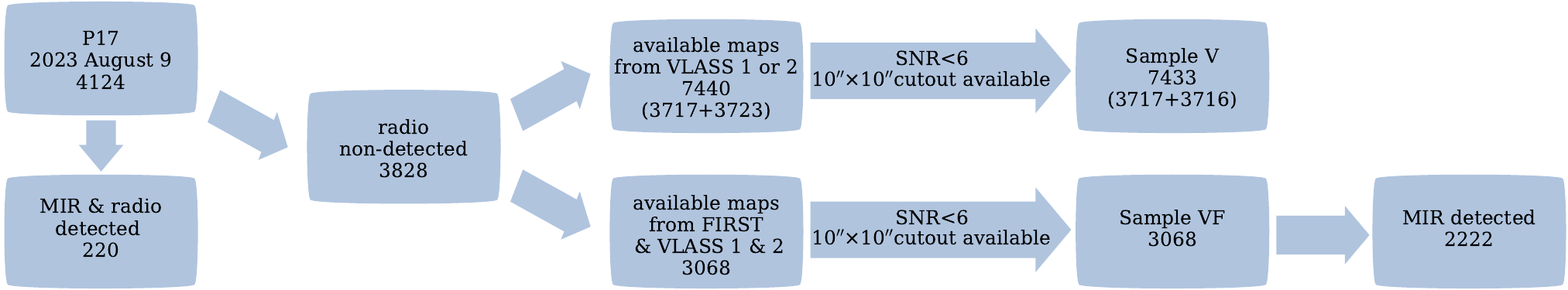}
 \caption{Steps of selecting Sample~V, Sample~VF, and the MIR subsamples by utilising the list of high-redshift AGNs collected in our catalogue P17. }\label{fig:sample}
\end{figure*}

To select suitable fields for the stacking analysis, we utilised the 2023 August 9 version of the P17 catalogue. This version contained $4124$ AGNs with redshifts equal to or exceeding 4. We discarded all $296$ objects which had been detected in either of the three radio surveys (FIRST, NVSS, VLASS), or had other publicly reported radio flux density measurement at any radio waveband, resulting in a list of $3828$ radio-non-detected AGNs. 

The coordinate list of these sources was used to collect the available empty-field `Quick Look' radio maps from the first and second epochs of the VLASS observing campaigns. The list of available images were obtained from the Canadian Astronomy Data Centre\footnote{\url{https://www.cadc-ccda.hia-iha.nrc-cnrc.gc.ca/en/}} (CADC) via a self-written \textsc{python} script, using the \textsc{astroquery} package \citep{2019AJ....157...98G}. We found empty-field radio maps for $3717$ and $3723$ AGNs from the first- and second-epoch VLASS campaigns, respectively. The downloaded maps were then investigated: we automatically discarded from the subsequent stacking analysis images with signal-to-noise ratios (SNR) higher than or equal to $6$ (these individual cases will be discussed in Sect.~\ref{new-detections}), and only maps with a complete $10\arcsec\times10\arcsec$ area available around the position of the given AGN were kept.  To obtain reliable noise estimates, we calculated the root mean square (rms) noise levels in a $10\arcsec\times10\arcsec$ square field around the image centre, with the central $2\farcs5\times2\farcs5$ area of the map masked out. This resulted in a final list of $7433$ maps, with $3717$ and $3716$ images at the first and second epochs, respectively. We will refer to the VLASS-only selected AGN list of the $7433$ maps as Sample~V. AGNs from Sample~V were also sorted into four redshift bins, with approximately equal number of sources ($\sim930$) in each, for both epochs (Fig.~\ref{fig:binning}). We performed  median stacking on the first- and second-epoch maps, both individually and the two epochs combined. We divided Sample~V into four bins, with limiting redshift values of $4.00\le z_1<4.14$, $4.14\le z_2<4.44$, $4.44\le z_3 <4.93$, and $4.93\le z_4<10.07$ (Fig.~\ref{fig:binning}). We note that the median value of the fourth redshift bin is $z_\mathrm{4,median}=5.66$, with only two objects above redshift $7.64$.

For a clear and unbiased comparison with stacking results using FIRST survey maps, another sample was defined, containing a list of AGNs with positions covered in both the FIRST and VLASS, with no radio detection in either of the two surveys. The AGN positions covered at both VLASS epochs were kept only. This yielded a list of $3068$ objects. As there were no radio maps with $\mathrm{SNR}\ge6$, the final sample consisted of empty-field maps for $3068$ AGNs for both of the VLASS observational epochs and FIRST, defining Sample~VF. AGNs in Sample~VF have redshifts in the range $4.00\le z \le 7.54$. The coordinates of the $3068$ AGNs in Sample~VF were cross-matched with the AllWISE Source Catalog\footnote{\url{https://wise2.ipac.caltech.edu/docs/release/allwise/}} \citep{2014yCat.2328....0C} using the NASA/IPAC Infrared Science Archive\footnote{\url{https://irsa.ipac.caltech.edu}}, applying a search radius of $5\arcsec$. We found MIR counterparts for $2222$ objects, which were observed and detected by \textit{WISE} with $\ge3\sigma$ in either of the four colour bands. Out of the $296$ radio-AGNs from P17, $220$ sources were detected by \textit{WISE} with  $\ge3\sigma$ in either of the four bands.
We illustrate the steps of selecting the Samples~V, VF, and the MIR subsamples in Fig.~\ref{fig:sample}.

To check the reliability of our stacking method, we also created two sub-samples of `artificial' AGN positions by randomly adding or subtracting $13\arcsec$ to/from the right ascensions or declinations of the real AGNs. The $13\arcsec$ displacement was selected so that the already downloaded images could be utilised, while excluding the parts of the tiles already used for the stacking around the real AGN positions. The two $\sim3700$-size samples centred on these artificial positions each contained  $10\arcsec\times10\arcsec$ size image cutouts. We stacked these images by both the mean and median methods for the entire batch, and with binning the sample by the values of the original real AGN redshifts. The stacking resulted in SNRs below 3, which are systematically lower than those of the median-stacked maps of real AGN positions, for both the full sample and the redshift-binned subsamples. As SNR levels for the mean-stacked images of the redshift-binned data of the real Sample~V and the artificial control sample were found to be the same, we only present results of mean stacking on the full sample. For Sample~VF, we only carried out median stacking as it was previously found that SNR values for real and artificial AGN positions were very similar using mean image stacking of FIRST maps \citep{2019MNRAS.490.2542P}.

\begin{figure*}
 \centering
 \includegraphics[width=.8\linewidth]{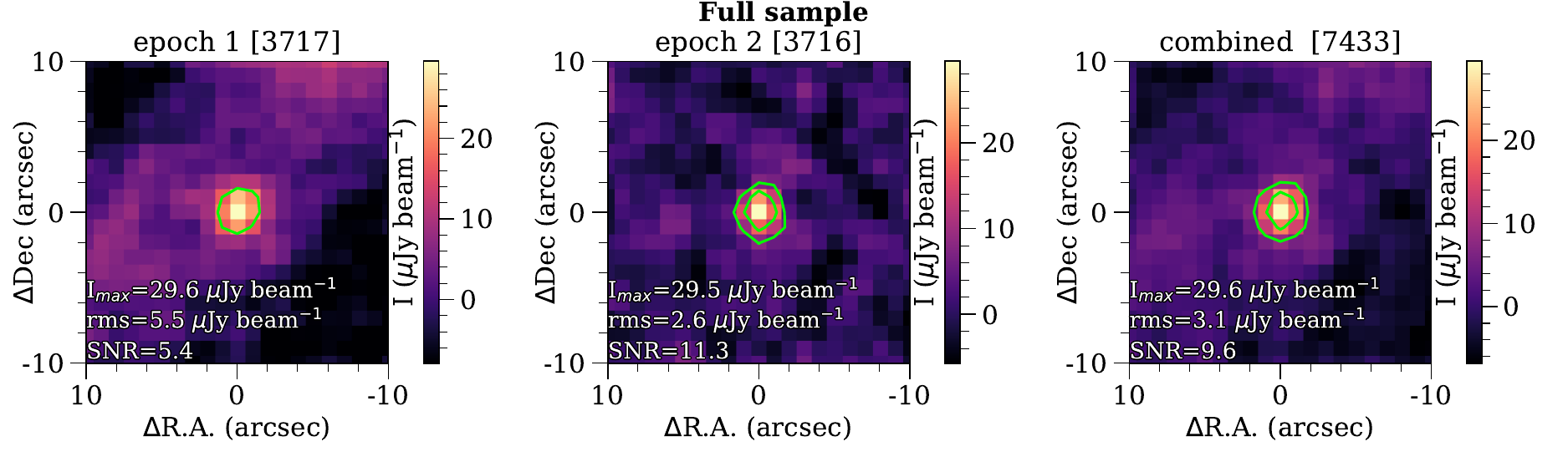}
 \caption{Mean stacked images centred on $3717$, $3716$, and $7433$ AGN positions (Sample~V) from the VLASS at epoch 1 and 2, as well as the combined data set, respectively. Intensity levels are denoted with green contour lines at $\pm3\sigma$ and $+6\sigma$ rms noise.}\label{fig:mean}
\end{figure*}

\begin{figure*}
 \centering
 \includegraphics[width=.8\linewidth]{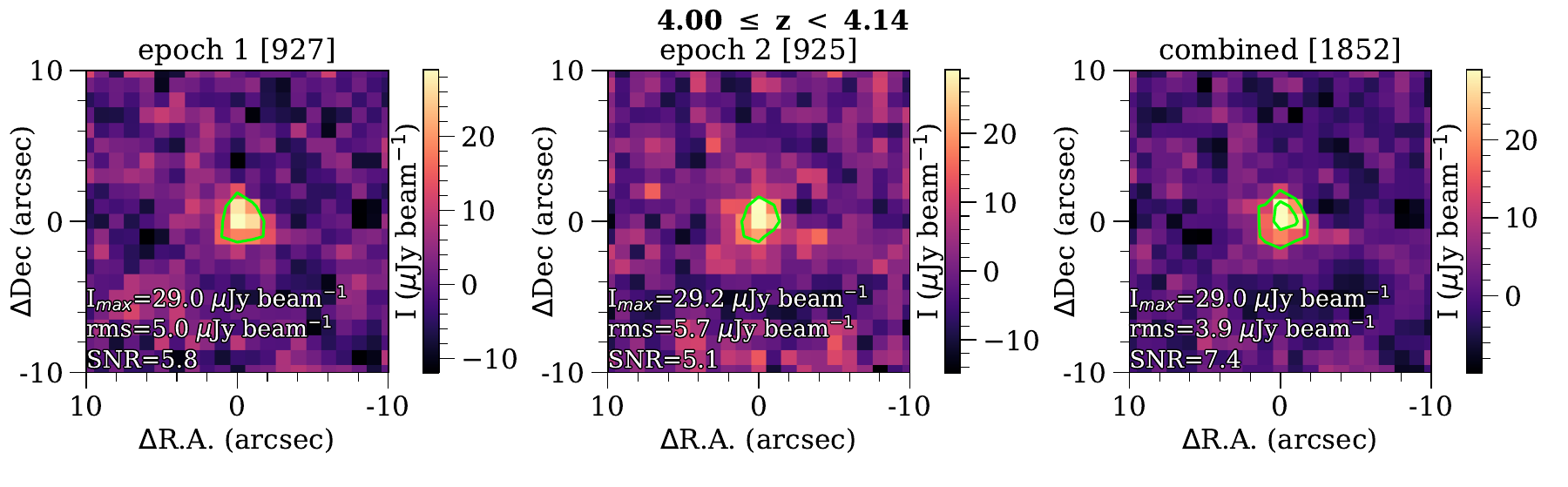}
 \includegraphics[width=.8\linewidth]{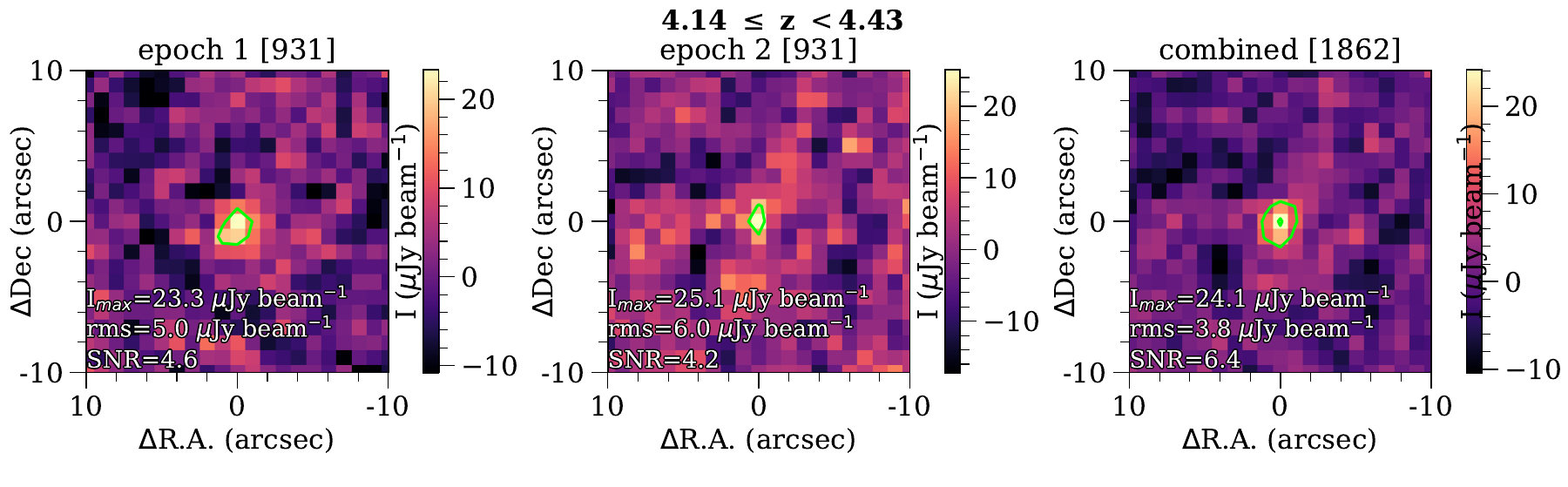}
 \includegraphics[width=.8\linewidth]{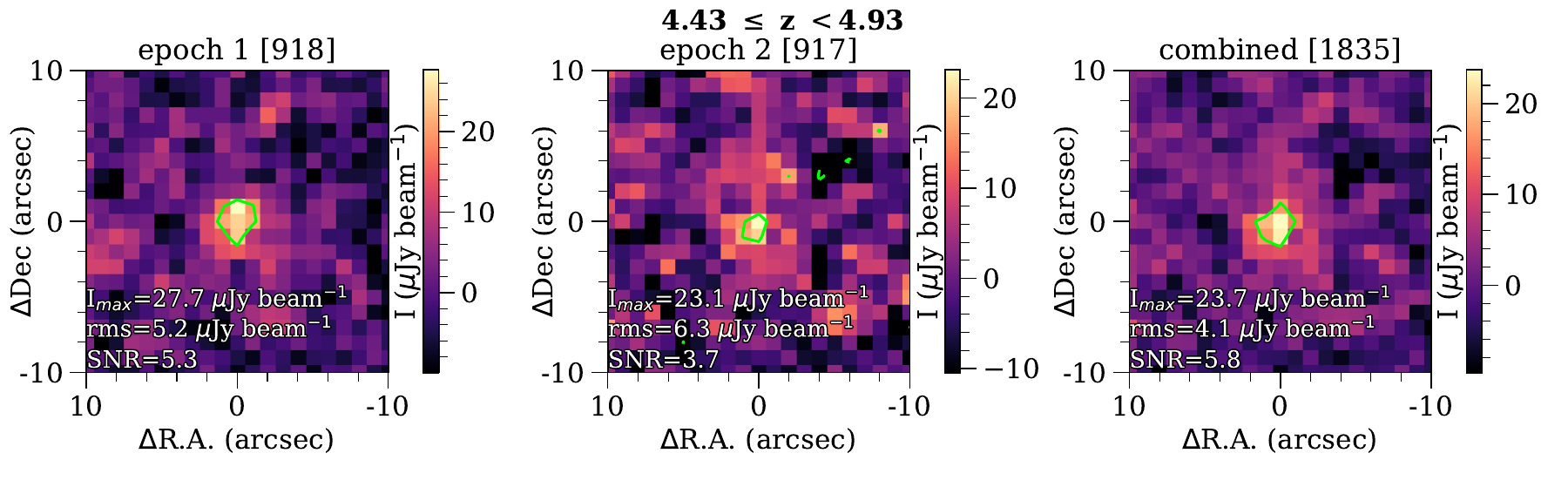}
 \includegraphics[width=.8\linewidth]{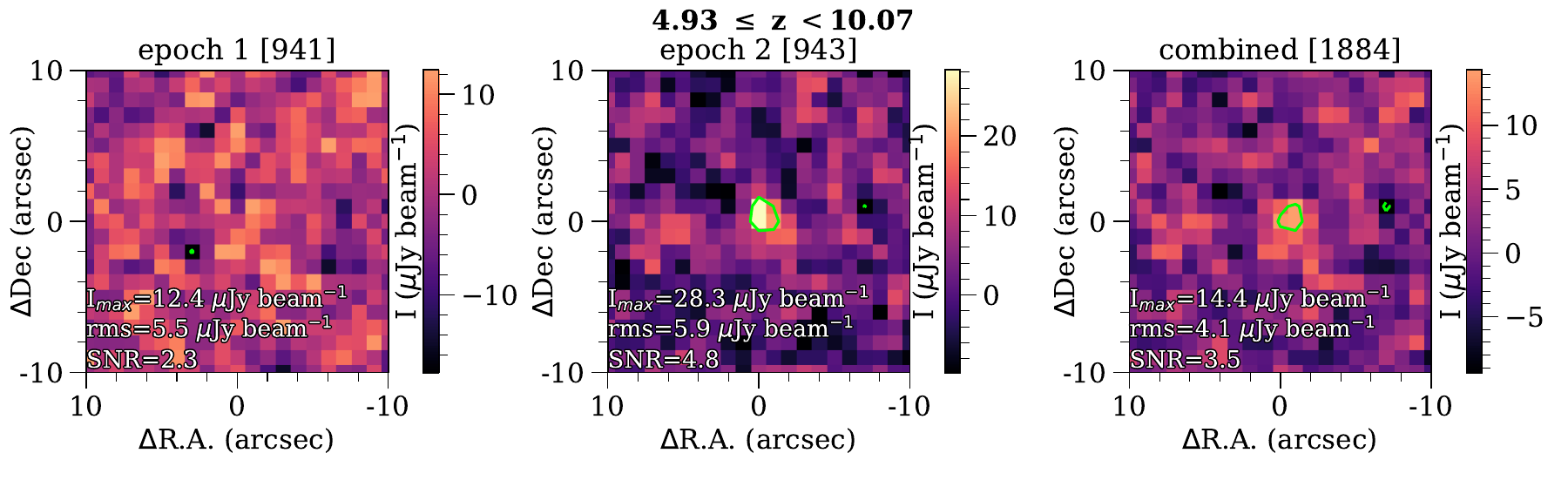}
 \includegraphics[width=.8\linewidth]{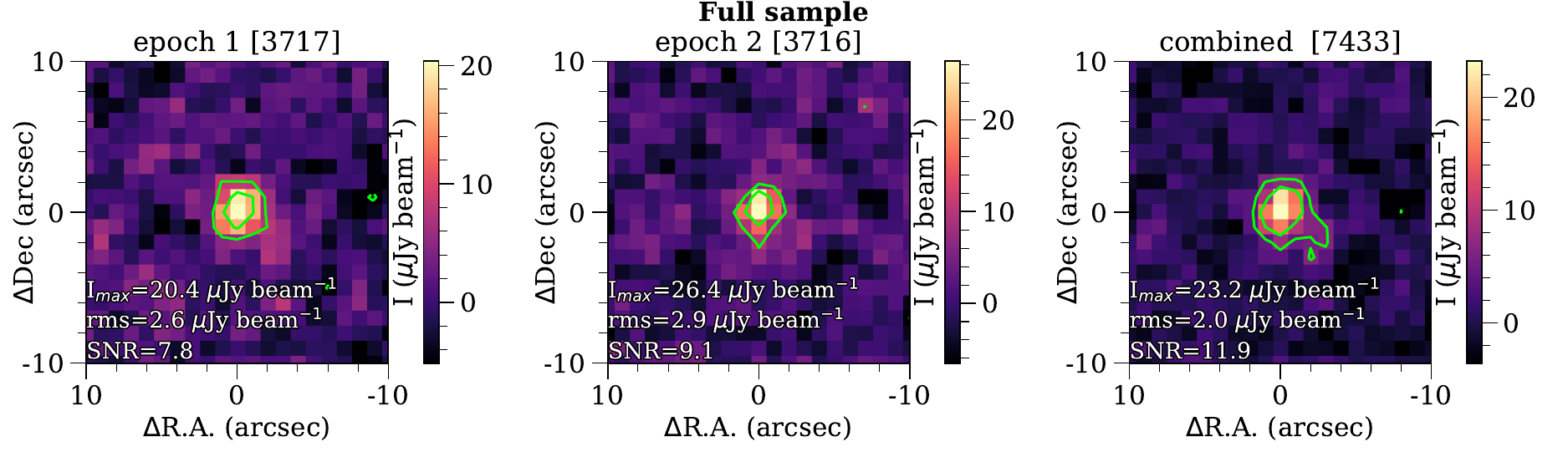}
 
 \caption{Median stacked images centred on $3717$, $3716$, and $7433$ AGN positions (Sample~V) from the VLASS at epoch 1 and 2, and the combined data set, respectively. Intensity levels are denoted with green contour lines at $\pm3\sigma$ and $+6\sigma$ rms noise.}\label{fig:median}
\end{figure*}

The images resulting from the stacking procedure (Figs.~\ref{fig:mean}, ~\ref{fig:median}, and \ref{fig:medianFV}) were then saved in \textsc{fits} format, and loaded into the US National Radio Astronomy Observatory (NRAO) Astronomical Image Processing System \citep[\textsc{aips},][]{2003ASSL..285..109G} package. To characterise the brightness distribution, we fitted each map with $\mathrm{SNR}\ge6$ with an elliptical Gaussian model component using the \textsc{jmfit} task. Flux density values for VLASS data were multiplied by a correction factor of $1.1$, following the VLASS Quick Look Images Users' Guide\footnote{\url{https://science.nrao.edu/vlass/data-access/vlass-epoch-1-quick-look-users-guide}}, to compensate for the $\sim10$ per cent systematic loss below $1$~Jy, while a factor of $1.4$ was applied as \textsc{clean} bias correction \citep[following][]{2007ApJ...654...99W} for the flux densities derived from stacked FIRST radio maps. Images with $\mathrm{SNR}<6$ were labeled as non-detection, and were assigned an upper limit on the flux densities as $6\sigma$ rms noise.

\section{Results}

\subsection{Mean and median stacking of VLASS maps}\label{sec:results_1}

Properties of the stacked images of Sample~V and the results of the model fitting are listed in Table~\ref{tab:results}. We were able to detect sub-mJy emission with $\mathrm{SNR}\ge6$ in the combined two-epoch data of the first ($4.00\le z \le4.14$) and second ($4.14\le z\le4.43$) redshift bins, and the full sample data of both individual epochs and the combined data. The flux density values for the median stacked full sample are $39 \pm 7$~\ujy, $34 \pm 7$~\ujy, and $36 \pm 5$~\ujy~ for the first, second, and combined epoch median-stacked data, respectively. 

Flux density values found for the mean-stacked second-epoch and combined VLASS data sets  agree with those of the median stacked images within the uncertainties.  
Only the  first and second redshift bins of the combined VLASS data sets were found to have  $\mathrm{SNR}\ge6$ in median stacking, with the former having a flux density slightly higher than what was found for the full-sample data. This can be explained by the abundance of $z\sim4$ AGNs in the sample dominating the population.

The ratio of peak intensity (measured in \ujb) and the fitted flux density (measured in \ujy), denoted here by $C$, is characteristic to the compactness of the emission \citep[e.g.][]{2014A&A...569A..35G,2018A&A...614A..87B,2020ApJ...896..117B,2020MNRAS.498.1278C}. For an unresolved source confined to within the restoring beam, one would expect $C=1$~beam$^{-1}$. 
The calculated compactness values for those maps with $\mathrm{SNR}>6$ span the range $0.52$~beam$^{-1}<C<0.85$~beam$^{-1}$, with a median value of $0.66$~beam$^{-1}$. This implies that the radio emission is slightly extended with the finer angular resolution of the VLASS. This is also indicated by the results of the Gaussian model fitting to the features seen in the stacked images in \textsc{aips}. The resolved nature of the sources could also explain the SNRs approaching but not reaching our somewhat conservative detection limit of 6, although there is a $70$ per cent increase in the number of AGNs involved in the stacking compared to our previous analysis of FIRST radio maps \citep{2019MNRAS.490.2542P}.

\begin{figure}
 \centering
 \includegraphics[width=\linewidth]{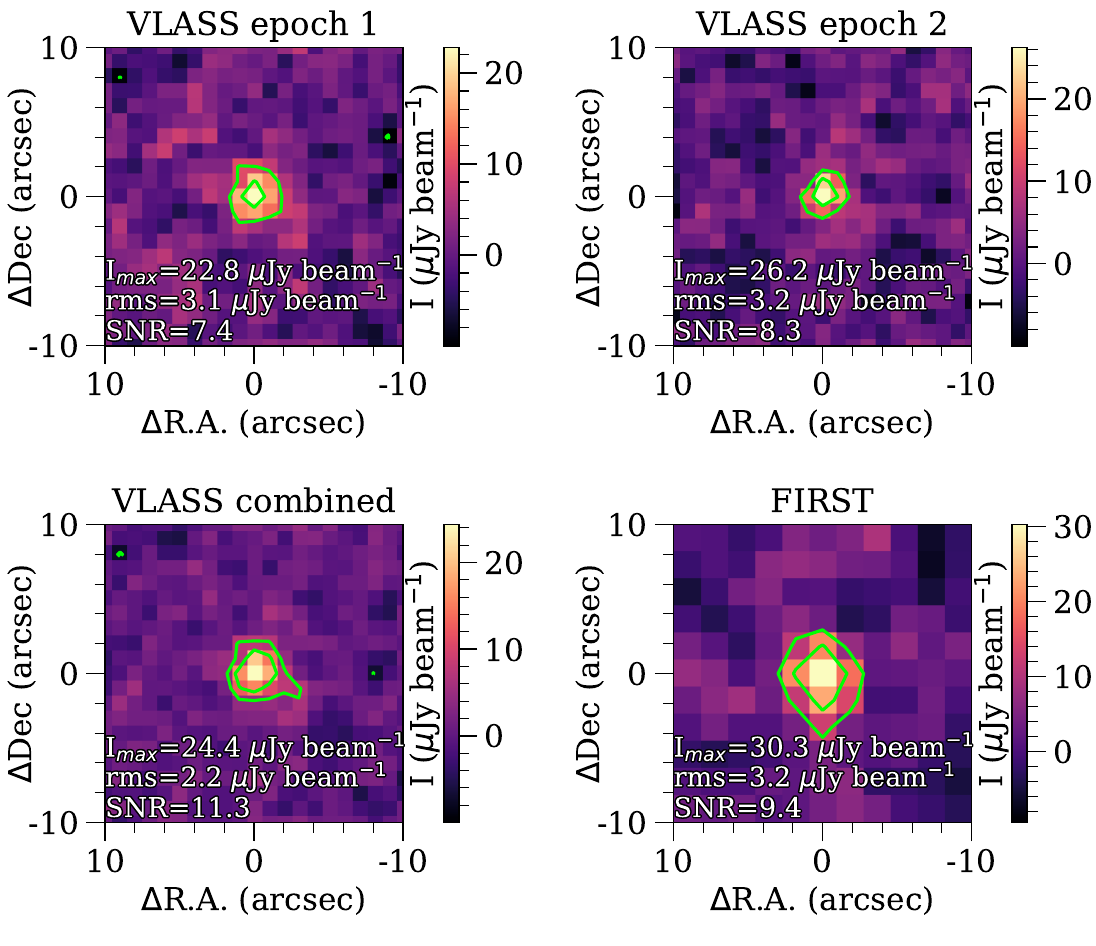}
 
 \caption{Median stacked images of the stacked FIRST and VLASS maps centred at $3068$ AGN positions (Sample~VF). The intensity levels are denoted with green contour lines at $\pm3\sigma$ and $+6\sigma$ rms noise.}\label{fig:medianFV}
\end{figure}

\begin{table*}
 \caption{Image properties and derived quantities of the mean and median stacked images centred on $3717$, $3716$, and $7433$ AGN positions from the VLASS at epoch 1 and 2, and the combined data set.}\label{tab:results}
 \centering\footnotesize
 \begin{tabular}{ccccccccccccccccccccc}
\hline\hline
&\multicolumn{6}{c}{Epoch 1}	& &	\multicolumn{6}{c}{Epoch 2}		& &\multicolumn{6}{c}{Combined}\\							
bin	&	$I_\mathrm{max} $	&	rms 	&	SNR	&	$S$	&$\sigma_S$	& $P_\mathrm{char} $ &&	$I_\mathrm{max} $	&	rms 	&	SNR	&	$S$&$\sigma_S$	&		 $P_\mathrm{char}$ &&	$I_\mathrm{max} $	&	rms 	&	SNR	&	$S$	&$\sigma_S$	& $P_\mathrm{char}$\\

\hline
\multicolumn{21}{c}{\textsc{mean}}\\\hline

all & 30 & 5 & 5.4 & <33 &  & <10 &  & 30 & 3 & 11.3 & 34 & 6& 2$-$10 &  & 30 & 3 & 9.6 & 42 & 8& 2$-$13 \\
\hline
\multicolumn{21}{c}{\textsc{median}}\\\hline
1 & 29 & 5 & 5.8 & <30 && <2 &  & 29 & 6 & 5.1 & <34 && <2 &  & 29 & 4 & 7.4 & 44 &9& 3 \\
2 & 23 & 5 & 4.6 & <30 && <2 &  & 25 & 6 & 4.2 & <36 && <3 &  & 24 & 4 & 6.4 & 29 &8& 2 \\
3 & 28 & 5 & 5.3 & <31 && <3 &  & 23 & 6 & 3.7 & <38 && <3 &  & 24 & 4 & 5.8 & <24 && <2 \\
4 & 12 & 5 & 2.3 & <33 && <10 &  & 28 & 6 & 4.8 & <35 && <11 &  & 14 & 4 & 3.5 & <25 && <8 \\
all & 20 & 3 & 7.8 & 39 &7& 2$-$12 &  & 26 & 3 & 9.1 & 34 &6& 2$-$11 &  & 23 & 2 & 11.9 & 36 &5& 2$-$11 \\\hline
\end{tabular} 
{\\\textit{Note.} Column 1 -- redshift bin (see Fig~\ref{fig:binning}), Columns 2, 8, 14 -- peak intensity of the stacked image in \ujb, Columns 3, 9, 15 --  rms noise level of the stacked image in \ujb, Columns 4, 10, 16 -- signal-to-noise ratio, Columns 5, 11, 17 -- flux density of the fitted model component in \ujy, Columns 6, 12, 18 -- uncertainty of the flux density in \ujy, Columns 7, 13, 19 -- characteristic monochromatic radio powers in $10^{24}$~\whz.
For non-detections (i.e. where SNR$<6$), an upper limit is given for the flux densities at $6\sigma$ rms noise level of the given image. The monochromatic powers were calculated for the lower and upper limiting  values in the given redshift bin, using the spectral index of $\alpha^*=-0.30$  determined from flux densities of Sample~VF in Section~\ref{sec:vlassfirstsec}. }
\end{table*}
\setlength{\tabcolsep}{6pt}\normalsize

\begin{table*}
 \centering\caption{Image properties and derived quantities of the median-stacked FIRST and VLASS maps centred at $3068$ AGN positions (Sample~VF). }
 \label{tab:vlassfirst}
 \begin{tabular}{ccccccc}
 \hline\hline
 Map & $I_\mathrm{max}$ & rms & SNR & $S$ &$\sigma_S$ & $P_\mathrm{char}$ \\ 
 & (\ujb) & (\ujb) & &(\ujy)& (\ujy)&($10^{24}$~\whz) \\
 \hline
epoch 1  & 23 & 3 &7.4 & 41& 8 & 2$-$6\\ 
epoch 2  & 26 & 3 &8.3 & 31& 7 & 2$-$5\\ 
combined  & 24 & 2 &11.3 & 36& 5 & 2$-$5\\ 
FIRST  & 30 & 3 &9.4 & 45& 7 & 2$-$7\\ 
 \hline
 \end{tabular}
 {\\\textit{Note.} The maximum intensity is determined for the original images, while the flux densities are modified with the correction factors. The minimum and maximum values for the characteristic powers are calculated using the minimum and maximum redshift values ($z_\mathrm{min}=4$, $z_\mathrm{max}=7.54$) in the sample, with spectral indices of the median value determined for the stacked image flux densities in this sample ($\alpha^*=-0.30$).}
\end{table*}

\subsection{Comparison with median-stacked FIRST radio maps}\label{sec:vlassfirstsec}

Results of the median stacking of VLASS and FIRST images centred on the $3068$ AGN positions in Sample~VF are shown in Fig.~\ref{fig:medianFV} and listed in Table~\ref{tab:vlassfirst}. The rms noise for the stacked FIRST maps is similar to the values for the single-epoch VLASS data sets with the same number of images in the procedure ($3068$).
The Gaussian model component fitting resulted in flux densities of $40 \pm 8$~\ujy, $31 \pm 6$~\ujy, and $35 \pm 5$~\ujy for the radio maps from the first, second, and combined epoch observations, respectively, while we found the value $44 \pm 6$~\ujy for the stacked images from the FIRST survey. The VLASS flux densities are similar to those found for the full sample (Sample~V) data. Although the peak intensity ($I_\mathrm{max,F}=30\pm9$~\ujb)
and the flux density ($S_\mathrm{F}=45\pm7$~\ujy) of the median-stacked FIRST images of Sample~VF are slightly lower than the values found in our previous FIRST stacking analysis, they are in agreement considering uncertainties \citep[$I_\mathrm{max,F}=35\pm11$~\ujb, $S_\mathrm{F}=52\pm16$~\ujy,][]{2019MNRAS.490.2542P}.

\section{Discussion}
\subsection{Signal-to-noise ratios of the stacked images}

Mostly because of the larger sample size \citep[c.f.][]{2019MNRAS.490.2542P}, the SNR values for most of the stacked VLASS images reach $\sim4-5$ (Table~\ref{tab:results}). We expect that future discoveries of new high-redshift AGNs, as well as the inclusion of maps from the upcoming third epoch of VLASS observations will allow us to reach the rms level of $\sim3$~\ujb in the binned data, and thus exceed the $6\sigma$ detection limit for all redshift bins. This would allow us to obtain information on the redshift dependence of the sub-mJy radio AGN population at $z\ge4$. There are currently $\sim300$ AGNs in P17 at redshift $z\ge6$, of which $279$ were included in the stacking analysis. As foreseen \citep{2019A&A...631A..85E,2023arXiv230812278T}, future deep optical--near infrared observations of the Vera C. Rubin Observatory combined with either the \textit{Euclid} or the Nancy Grace Roman Space Telescopes could result in the discovery of $\sim200-250$ new distant AGNs at $6<z<9$. An increase of this magnitude in the number of spectroscopically identified $z\ge6$ AGNs might enhance the chance of recovering faint radio emission from the radio-quiet population at the highest redshifts.

\subsection{Radio spectrum of the sub-mJy sample}

The median value of the spectral indices ($\alpha$, using the convention of $S_\nu\sim\nu^\alpha$) calculated for those AGNs form P17 \citep{2017FrASS...4....9P} which were individually detected in both the VLASS and FIRST surveys ($S\gtrsim1$~mJy) is $\alpha=-0.40\pm0.05$, which is in good agreement with the median value determined for 500\,000 VLASS- and FIRST-detected sources \citep[$\alpha=-0.39$,][]{2021ApJS..255...30G}. 

Calculating the spectral index using the flux densities form the models fitted to the median stacked images of Sample~VF (Section~\ref{sec:vlassfirstsec}, Table~\ref{tab:vlassfirst}), the characteristic spectral properties of radio-detected and non-detected AGNs can be compared.  We found that the two-point spectral index for our stacked sub-mJy sample is $\alpha^*=-0.30\pm0.15$, which is in agreement with that of the radio-detected population. This implies that the overall spectral properties of `average' radio-emitting AGNs are similar, in spite of their power, supporting the idea of a continuous radio luminosity function (LF) rather than the bimodal distribution of radio-loud and radio-quiet objects. The possibility of a continuous transition between the SF and AGN dominance in the overall radio emission detected from the sources might also support this notion \citep[e.g.][]{2021MNRAS.506.5888M}. The flat characteristic spectrum implies that for the majority of the sub-mJy sources, the radio emission originates at least partially from AGN activity.

Studies of the spectral properties of samples of $\sim5500$ star-forming galaxies (SFGs) in the MIGHTEE-COSMOS and ELAIS-N1 fields found that radio-detected SFGs usually show a steeper spectrum between $1-5$~GHz   \citep{2021MNRAS.507.2643A, 2023arXiv230306941A}. Similarly, in the framework of the VLA-COSMOS 3 GHz Large Project,  \citet{2017A&A...602A...4D} found that the median spectral index of the SFG population was consistent with $\alpha_\mathrm{1.4\,GHz}^\mathrm{3\,GHz}=-0.8$ at redshifts $z>2$. The flat radio spectrum found for our Sample~VF AGNs suggests that the sub-mJy level radio emission is dominated by compact AGN jets, with less significant contribution from star formation in the host galaxies, steep-spectrum jet or lobe emission, or AGN-driven winds \citep[e.g][]{2018MNRAS.477..830H, 2019MNRAS.485.2710J,2020MNRAS.499.5826S}.

\subsection{Characteristic radio powers}
We calculated the characteristic monochromatic radio powers for all stacked images, following the expression of
\begin{equation}
    P_\mathrm{char}=4\pi d_\mathrm{L}^2 S (1+z)^{-\alpha-1},
\end{equation}
where $S$ is the bias-corrected flux density value, $d_\mathrm{L}$ the luminosity distance at redshift $z$, and $\alpha$ the radio spectral index. We chose the lowest and highest redshift values in each redshift bin to calculate the possible range of radio powers, using the characteristic spectral index determined from the stacked images of the $3068$ sources in Sample~VF ($\alpha^*=-0.30$). Radio powers corresponding to the lower and upper limiting redshift values are listed in Tables~\ref{tab:results} and \ref{tab:vlassfirst}, for the AGN positions covered by the VLASS (Sample~V) and the VLASS--FIRST overlapping regions (Sample~VF), respectively. We found that the characteristic monochromatic radio powers for both the $1.4$ and $3$~GHz data are in the order of $10^{24}$~\whz. This implies that AGNs in the stacked sample, even though they are close to the faint end of the LF, generally exceed the radio powers of $10^{22-23}$~\whz characteristic for star formation in starburst galaxies \citep[e.g.][]{1991ApJ...378...65C,2006ApJ...652..177K,2012MNRAS.423.1325A,2013ApJ...768...37C}, thus belong to the intermediate-luminosity population \citep[e.g.][]{2005MNRAS.362...25B,2017A&A...602A...6S,2023MNRAS.519.4902S} at both frequency bands. 
The flat spectrum and high characteristic radio powers ($P\sim10^{24}-10^{25}$~\whz) suggest that the dominant radio emission mechanism in the sub-mJy sources is AGN activity \citep[e.g.][]{2013ApJ...768...37C,2021MNRAS.503.1780J}.

\subsection{Mid-infrared colours}
\begin{figure}
     \centering
 \includegraphics[width=\linewidth]{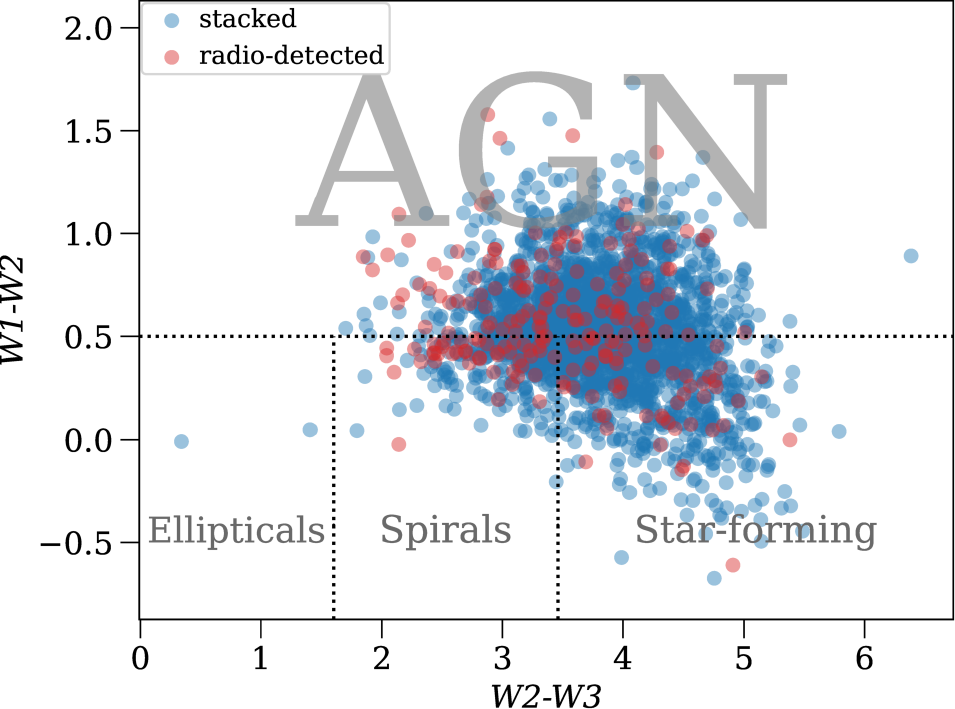}
  \caption{Three-band \textit{WISE} colour--colour diagram with $\mathrm{SNR}>3$ detections from the \textit{AllWISE} data release. Blue and red circles denote the $2222$ and $220$ AGNs from Sample~VF (radio non-detected) and P17 (radio-detected), respectively. The magnitudes are \textit{WISE} catalogue values in the Vega system. The sub-class limits are applied from e.g. \citet{2016MNRAS.462.2631M} and \citet{2018MNRAS.480..358W}.}
  \label{fig:wisecolours}
\end{figure}

Following the findings of e.g. \citet{2016MNRAS.462.2631M, 2018MNRAS.480..358W}, we compared the  $W2-W3$ vs. $W1-W2$ colours of the $220$ radio- and MIR-detected AGNs from P17 and the $2222$ MIR-detected objects from Sample~VF (Fig.~\ref{fig:wisecolours}). The $W1-W2>0.5$~mag colour cut categorises $1167$ objects as AGNs, while $801$ reside in the star-forming section defined by $W1-W2<0.5$~mag and $W2-W3>3.46$~mag. For radio-detected AGNs from P17, we found that $118$ objects lie in the AGN region and $56$ show characteristic MIR colours of SFGs. Thus, the AGN fraction is very similar, $\sim54$ and $\sim52$ per cent for the radio-detected and radio-nondetected (stacked) sample, respectively, hinting that the AGN-related radio emission is also similarly distributed, implying that the sub-mJy radio flux densities are still predominantly originating from AGN activity.

\subsection{New bona fide radio sources}
\label{new-detections}

\begin{figure*}
    \centering
    \includegraphics[width=0.4\linewidth]{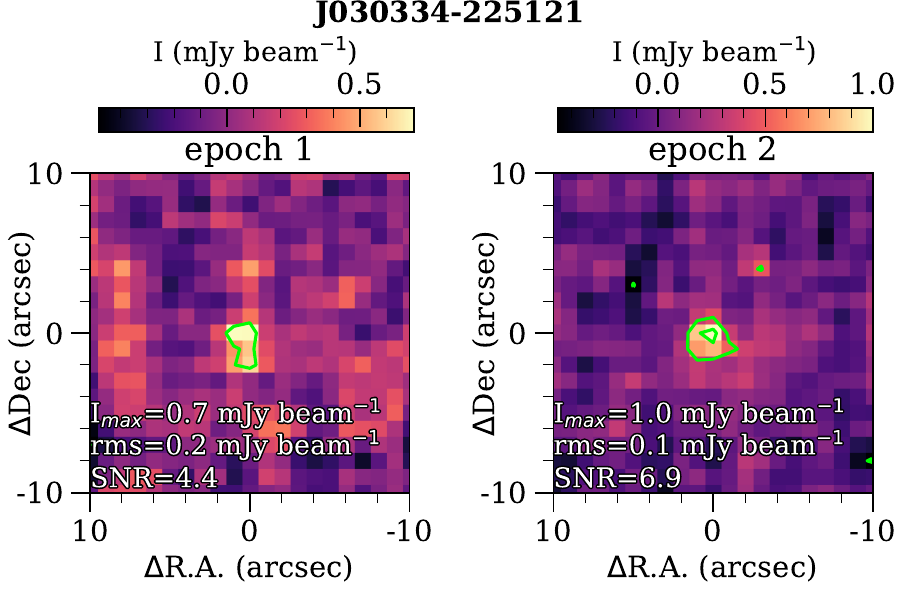}~~~~
    \includegraphics[width=0.4\linewidth]{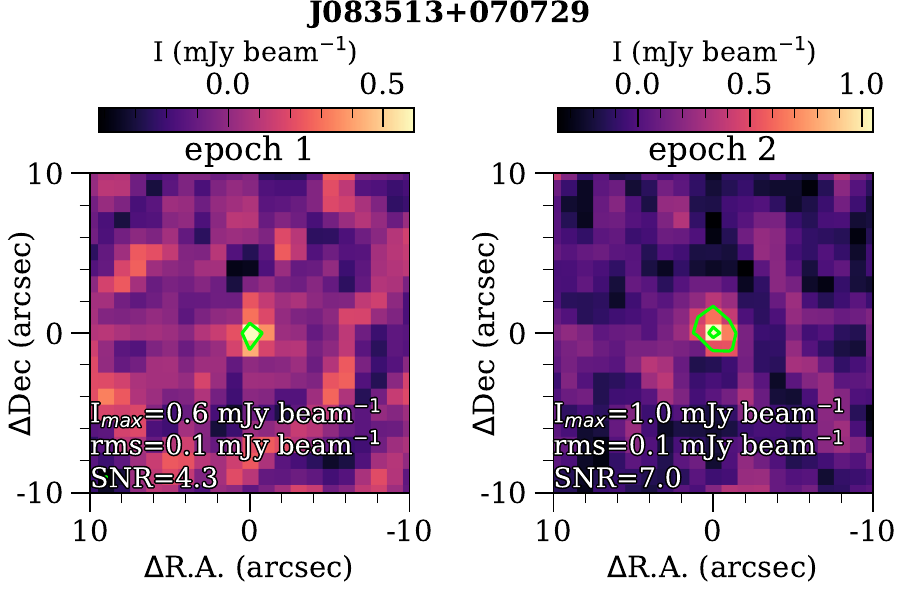}\\
    \includegraphics[width=0.4\linewidth]{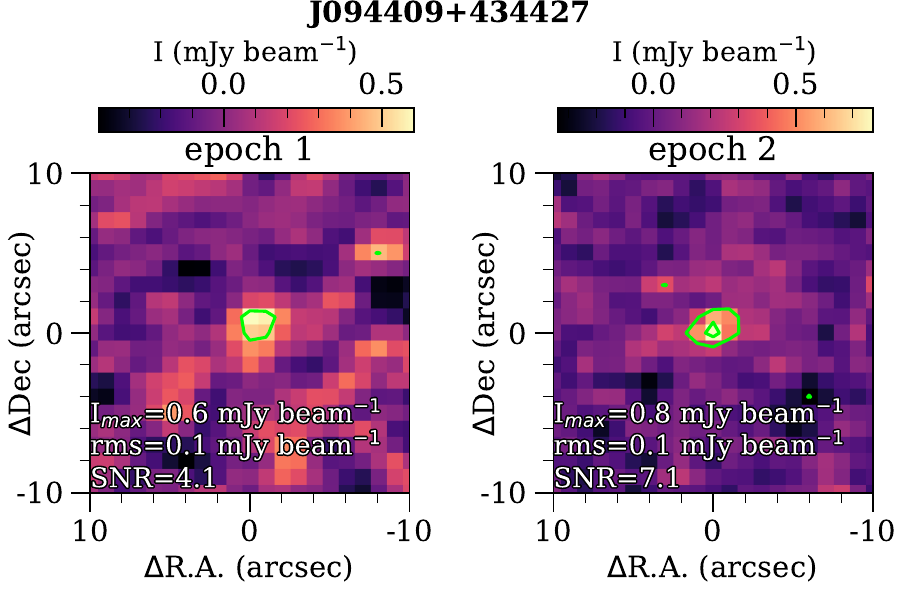}~~~~
    \includegraphics[width=0.4\linewidth]{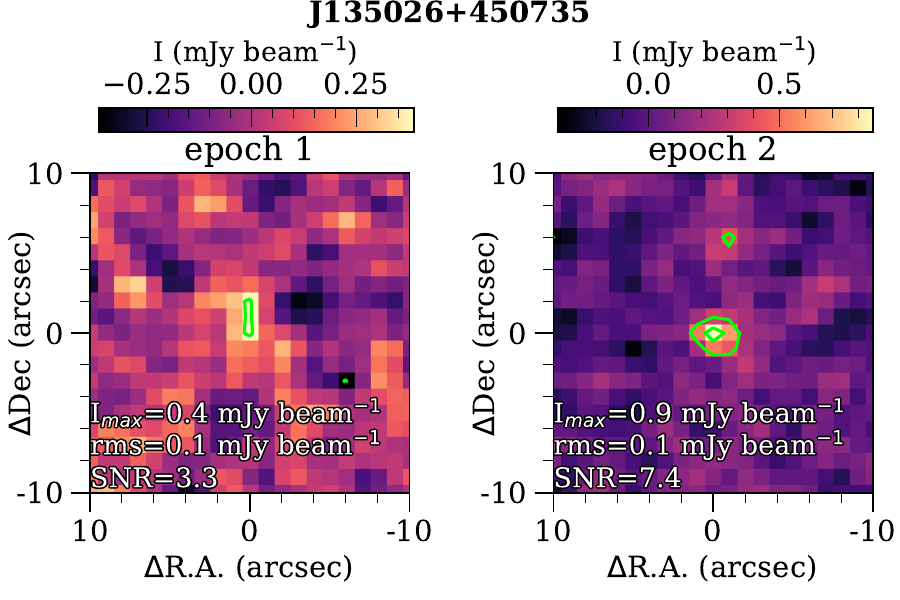}\\
    \includegraphics[width=0.4\linewidth]{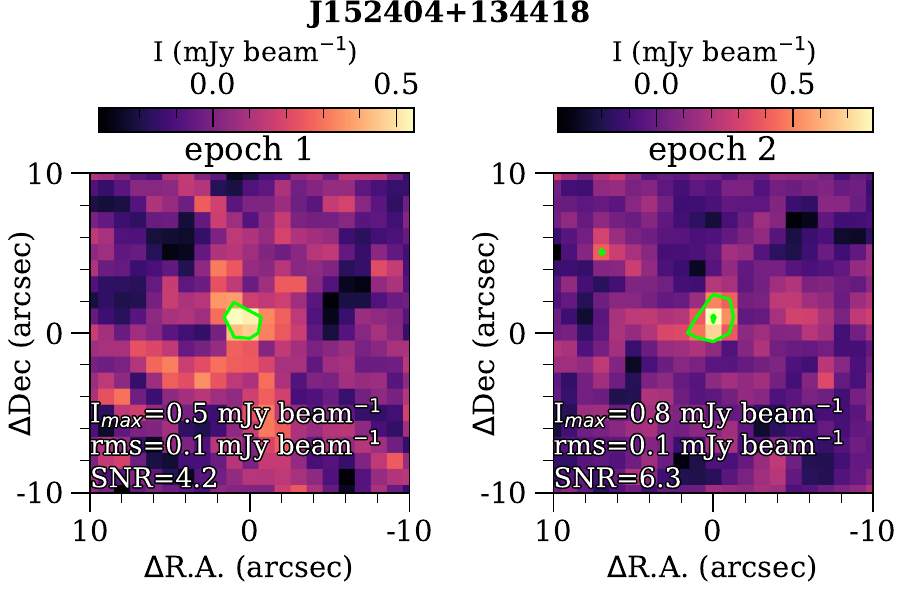}~~~~
    \includegraphics[width=0.4\linewidth]{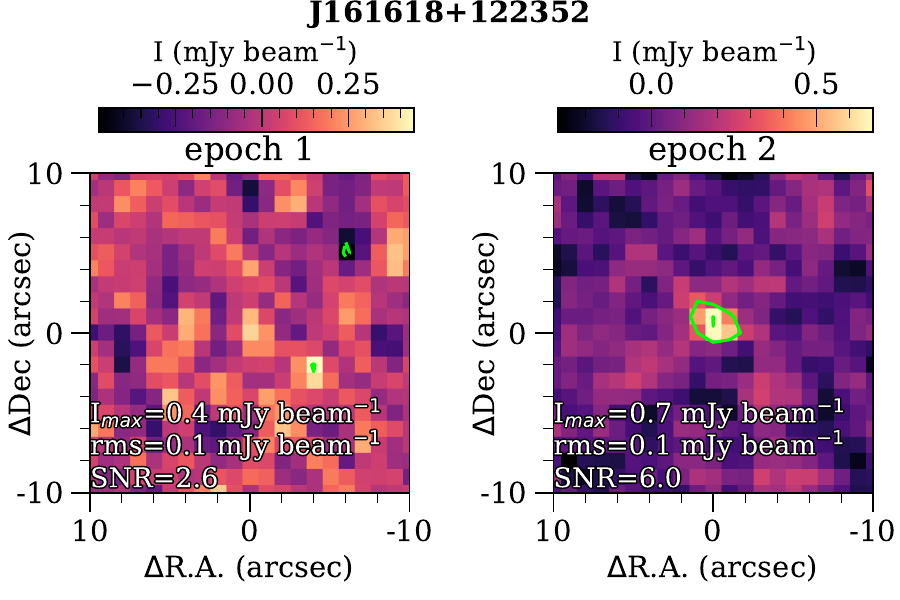}
    \caption{First- and second-epoch VLASS image cutouts of AGNs with no radio counterpart in P17 that showed $\mathrm{SNR}\ge6$ in the second-epoch VLASS observations. Intensity levels are denoted with green contour lines at $\pm3\sigma$ and $+6\sigma$ rms noise.} \label{fig:snrgrt6}
\end{figure*}

During the sample selection, we disregarded from empty-field radio maps those with $\mathrm{SNR}\ge6$. No such image cutouts were excluded from the first-epoch VLASS analysis, while there were $6$ AGNs found in the second-epoch VLASS data which showed evidence for significant radio emission (Fig.~\ref{fig:snrgrt6}, Table~\ref{tab:snrgrt6}). None of them were reported as radio sources in the first-epoch VLASS Quick Look catalogue \citep{2020RNAAS...4..175G}. We fitted elliptical Gaussian model components to these images in \textsc{aips} using the task \textsc{jmfit}, and  multiplied the flux densities with the correction factor of $1.1$, resulting in $\sim1$~mJy level radio emission for all of them (Table~\ref{tab:snrgrt6}). We also cross-matched this list of AGNs with the first version of the VLASS Quick Look epoch 2 source catalogue\footnote{\url{https://cirada.ca/vlasscatalogueql0}}. Only two of the AGNs in the list (marked with * in Table~\ref{tab:snrgrt6}) had counterparts in this catalogue: J030333.77$-$225121.6 ($S_\mathrm{VLASS}=1.56\pm0.33$~mJy) and J152404.22+134417.8 ($S_\mathrm{VLASS}=1.11\pm0.30$~mJy). Flux densities from our model fitting and from the VLASS Quick Look epoch 2 catalogue agree within the uncertainties. 
The remaining 4 sources listed in Table~\ref{tab:snrgrt6}  can be considered as newly identified high-redshift radio-emitting AGNs. Their detection only at the second epoch may be due to the improvement of the VLASS calibration pipeline (since they are close to the detection limit of the survey), and might be confirmed later when the higher quality single-epoch images will be available. Another possibility is that these AGNs brightened up during $\sim3$~yr between the two epochs of VLASS observations \citep[e.g.][]{2020ApJ...905...74N,2021IAUS..359...27N}.

\begin{table}
     \centering
     \caption{AGNs with no known radio counterpart in P17 that showed $\mathrm{SNR}\ge6$ in the second-epoch VLASS observations.}
    \label{tab:snrgrt6}
    \begin{tabular}{lcccc}
    \hline
    \hline
    Name & R.A. & Dec. & $z$ & $S$ \\
     & ($\degr$) & ($\degr$) & & (mJy) \\
    \hline
    J030334$-$225121$^*$ & 45.890750 & $-22.855944$ & 4.57 & $1.62 \pm 0.36 $ \\
    J083513+070729 & 128.804804 & 7.124819 & 4.05 & $1.12 \pm 0.18 $ \\
    J094409+434427 & 146.036763 & 43.740848 & 4.56 & $1.04 \pm 0.25 $ \\
    J135026+450735 & 207.607058 & 45.126289 & 4.12 & $0.84 \pm 0.21 $ \\
    J152404+134418$^*$ & 231.017631 & 13.738205 & 4.79 & $1.09 \pm 0.28 $ \\
    J161618+122352 & 244.075339 & 12.397644 & 4.29 & $0.84 \pm 0.23 $ \\
    \hline
    \end{tabular}
 {\\\textit{Note.} Sources marked with $^*$ are found in the VLASS Quick Look epoch 2 catalogue.}  
\end{table}
 
\section{Summary and conclusions}
We selected two samples (V and VF) of high-redshift ($z\ge4$) active galactic nuclei individually non-detected in radio from our catalogue \citep[P17,][]{2017FrASS...4....9P} to perform image stacking, utilising empty-field radio maps from the VLASS and FIRST surveys at $3$ and $1.4$~GHz, respectively, centred on the positions of the AGNs. Mean and median stacking were carried out using the VLASS-only images from Sample~V with full-sample and four redshift-binned data sets. We found that the $70$ per cent increase of source numbers compared to the sample used in our previous stacking study with FIRST-only images \citep{2019MNRAS.490.2542P}, and the utilisation of two-epoch VLASS observations resulted in signal-to-noise ratios approaching the $6\sigma$ detection threshold, strengthening the expectations that the inclusion of yet-to-be-completed third-epoch VLASS data, combined with future deep optical--near infrared observations (e.g. with the Vera C. Rubin Observatory, \textit{Euclid}, or the Nancy Grace Roman Space Telescopes) will allow us to start investigating the redshift dependence of the sub-mJy radio AGN population. We found characteristic radio powers in the order of $10^{24}-10^{25}$~\whz, which is in agreement with those of the intermediate-luminosity radio AGNs previously found in FIRST-stacked images \citep{2019MNRAS.490.2542P}. Using the $3068$ AGN images from Sample~VF (i.e. for AGN positions covered in both VLASS and FIRST), we found that the sub-mJy population has a flat characteristic radio spectrum between $1.4$ and $3$~GHz, which is in agreement with the value determined for high-redshift radio-detected AGNs from P17. The flat spectrum implies that the sub-mJy radio emission predominantly originates from AGN activity, rather than from star formation. The investigation of MIR photometric data from the \textit{AllWISE} data release also supports this finding as the majority of the MIR-detected sources show characteristic colours of AGNs in the three-band colour--colour diagram.

We also report four new radio-emitting AGNs at $4 < z < 4.6$ found in the second-epoch VLASS images with $\mathrm{SNR}\ge6$ detection and mJy-level flux densities.

\section*{Acknowledgements}
The authors thank the anonymous reviewer for their insightful suggestions that led to the improvement of the manuscript.

The National Radio Astronomy Observatory is a facility of the National Science Foundation operated under cooperative agreement by Associated Universities, Inc. CIRADA is funded by a grant from the Canada Foundation for Innovation 2017 Innovation Fund (Project 35999), as well as by the Provinces of Ontario, British Columbia, Alberta, Manitoba and Quebec. 

This publication makes use of data products from the \textit{Wide-field Infrared Survey Explorer}, which is a joint project of the University of California, Los Angeles, and the Jet Propulsion Laboratory / California Institute of Technology, funded by the National Aeronautics and Space Administration. 

This research has made use of the NASA/IPAC Infrared Science Archive, which is funded by the National Aeronautics and Space Administration and operated by the California Institute of Technology.

This work made use of \textsc{topcat}
\citep[Tool for OPerations on Catalogues And Tables,][]{2005ASPC..347...29T}, and the following \textsc{python} packages: \textsc{astropy} \citep{astropy:2013,astropy:2018,astropy:2022}, \textsc{astroquery}  \citep{2019AJ....157...98G}, \textsc{matplotlib}  \citep{Hunter:2007}, \textsc{numpy} \citep{harris2020array}, and \textsc{pandas} \citep{mckinney-proc-scipy-2010,reback2023pandas}.

This research was supported by the Hungarian National Research, Development and Innovation Office (NKFIH), grant number OTKA K134213. This project has received funding from the HUN-REN Hungarian Research Network.

\section*{Data Availability}
The latest version of the catalogue of high-redshift AGNs can be obtained from \url{https://staff.konkoly.hu/perger.krisztina/catalog.html} or from ViZieR (\url{https://cdsarc.cds.unistra.fr/viz-bin/cat/J/other/FrASS/4.9}). Earlier catalogue versions may be requested from the corresponding author upon reasonable request.
The lists of the objects used in the stacking analyses are provided as supplementary material in \textsc{csv} format. The empty-field radio maps were obtained from  the FIRST cutout service (\url{http://sundog.stsci.edu/index.html}) and the Canadian Astronomy Data Centre (\url{https://www.cadc-ccda.hia-iha.nrc-cnrc.gc.ca/}). Infrared data were collected using the NASA/IPAC Infrared Science Archive (\url{https://irsa.ipac.caltech.edu/}).

\bibliographystyle{mnras}
\bibliography{vlass_stack_bib} 






\bsp	
\label{lastpage}
\end{document}